\documentclass{elsart}
\usepackage{natbib}
\usepackage{epsfig}
\usepackage{amssymb}
\usepackage{amsmath}
\usepackage{color}
\usepackage{graphicx}

\begin{document}

\begin{frontmatter}

\title{The beaming of external Compton emission}
\author{Anne Hutter\corauthref{cor}}
\address{Astrophysical Institute of Potsdam, An der Sternwarte 16, 14482 Potsdam, Germany}
\corauth[cor]{Corresponding author}
\ead{ahutter@aip.de}

\author{Felix Spanier\thanksref{footnote2}}
\address{Lehrstuhl f\"ur Astronomie, University of W\"urzburg, Am Hubland, 97074 W\"urzburg, Germany}
\thanks[footnote2]{This work has been partially funded by the Deutsche Forschungsgemeinschaft (DFG) under contract SP 1124/3}

\begin{abstract}
We consider a relativistically moving blob consisting of an isotropic electron distribution 
that Compton-scatters photons from an external isotropic radiation field. 
We compute the resulting beaming pattern, i.e. the distribution of the scattered photons,
in the blob frame as well as in the observer's frame by using the full Klein-Nishina cross 
section and the exact incident photon distribution.
In the Thomson regime the comparison of our approach with \citet{Dermer1995} results in concurrent 
characteristics but different absolute number of the scattered photons by a factor of $f_{corr}=3.09$.
Additionally, our calculation yields a slightly lower boost factor which varies the more from 
the corresponding value in \citet{Dermer1995} the higher the spectral index $p$ of the electron 
distribution gets.
\end{abstract}


\end{frontmatter}

\parindent=0.5 cm

\section{Introduction}

Blazars are flat radio spectrum active galactic nuclei that are dominated by the nonthermal emissions 
of superluminal moving bulk plasma blobs emitted from the core \citep{Zensus1987}.
Their strong high energy emissions show a characteristic double bump structure in their spectra \citep{Sanders1989}. 
In leptonic models, the lower energy bump is caused by Synchrotron emission while the bump at higher energies is produced by inverse Compton-scattering. 
According to Synchrotron Self Compton models (SSC) Synchrotron photons are produced by accelerated particles in the jet, 
while gamma ray photons arise from the Inverse Compton-scattered Synchrotron photons \citep{Maraschi1992,Marscher1994}. 
However, external photons that are not produced inside the jet can also be scattered by the Inverse Compton effect and 
can contribute to the emission. 
External Compton models (EC) consider photons from the accretion disk or the Broad Line Region \citep{Blandford1993,Dermer1992,Dermer1993,Sikora1994} while 
conventional one zone SSC models consider the emission in the blob as homogeneous. These models assume
that the variability of blazar emissions is limited by the size of the blob. In contrast, 
two zone SSC models \citep{Weidinger2010} allow for shorter variabilities but only spatially resolved models are 
suitable for even shorter variabilities. 
They take angular dependent properties of the jet into account like the direction of magnetic fields and 
spatial variations of electron and photon densities inside the blob.

In this paper we consider spatially resolved SSC models with an EC component and are interested in ascertaining the angular radiation 
characteristics of the EC component. \cite{Dermer1995} assumed that this beaming pattern is produced by a relativistic moving blob 
consisting of isotropic distributed electrons that Thomson-scatter photons from an external isotropic radiation field.
As a result the isotropic external photons are more strongly boosted than isotropically distributed photons in a blob within the jet.
While Dermer's calculations are based on ultrarelativistic electrons and scattering in the Thomson regime, in this paper
we will consider all possible electron energies obeying the Thomson as well as the Klein-Nishina regime.
Using the exact Klein-Nishina cross section we compute the distribution of scattered photons numerically, examine the dependencies of physical parameters 
and determine the resulting intensity for an electron distribution given by a power law.
Such a calculation has already been executed by \cite{Georganopoulos2001} but in contrast to them we do not use the formula of the  scattered photons by \cite{Jones1968} that averages over the photons' emergent angles. In fact we will inherit the first steps of Jones exact calculation, i.e. not relying on the head on approximation, and continue using exact expressions.

We organize the remaining paper as follows. In Section \ref{sec_calc} the necessary steps are shown to calculate the scattered spectrum exactly. In Section \ref{sec_results} the results in the blob frame as well in the observer's frame are presented and compared to both \cite{Dermer1995} and \cite{Georganopoulos2001}. 

\section{Calculating the spectrum}
\label{sec_calc}
\begin{figure}
\centering
\begin{picture}(0,0)%
\includegraphics{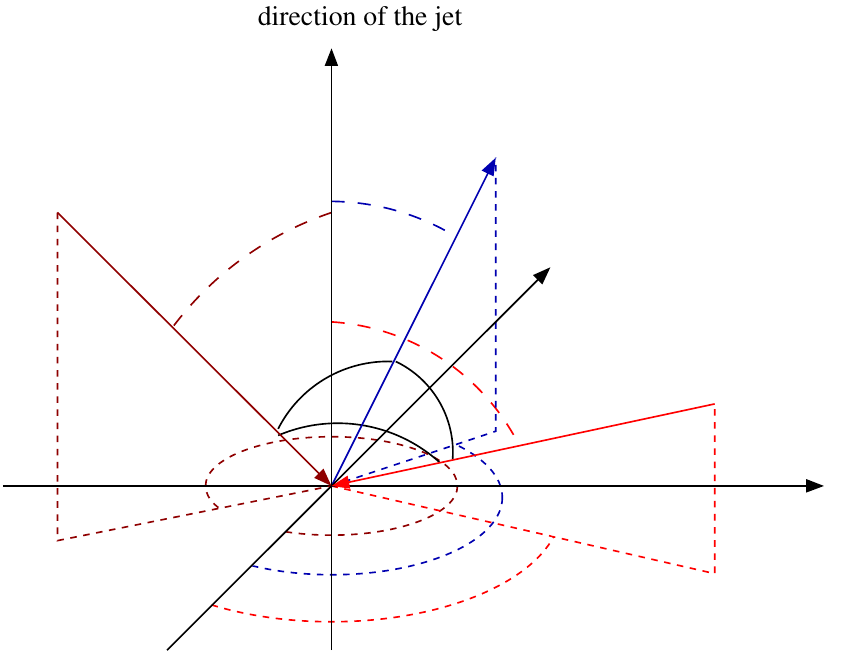}%
\end{picture}%
\setlength{\unitlength}{4144sp}%
\begingroup\makeatletter\ifx\SetFigFont\undefined%
\gdef\SetFigFont#1#2#3#4#5{%
  \reset@font\fontsize{#1}{#2pt}%
  \fontfamily{#3}\fontseries{#4}\fontshape{#5}%
  \selectfont}%
\fi\endgroup%
\begin{picture}(3926,2977)(-11,-8623)
\put(1167,-7251){\makebox(0,0)[lb]{\smash{{\SetFigFont{8}{9.6}{\rmdefault}{\mddefault}{\updefault}{\color[rgb]{.56,0,0}$\theta_{\alpha_1}$}%
}}}}
\put(1669,-6964){\makebox(0,0)[lb]{\smash{{\SetFigFont{8}{9.6}{\rmdefault}{\mddefault}{\updefault}{\color[rgb]{0,0,.69}$\theta_e$}%
}}}}
\put(1792,-8181){\makebox(0,0)[lb]{\smash{{\SetFigFont{8}{9.6}{\rmdefault}{\mddefault}{\updefault}{\color[rgb]{0,0,.69}$\phi_e$}%
}}}}
\put(1720,-8407){\makebox(0,0)[lb]{\smash{{\SetFigFont{8}{9.6}{\rmdefault}{\mddefault}{\updefault}{\color[rgb]{1,0,0}$\phi_{\alpha}$}%
}}}}
\put(1137,-7833){\makebox(0,0)[lb]{\smash{{\SetFigFont{8}{9.6}{\rmdefault}{\mddefault}{\updefault}{\color[rgb]{.56,0,0}$\phi_{\alpha_1}$}%
}}}}
\put(2089,-7629){\makebox(0,0)[lb]{\smash{{\SetFigFont{8}{9.6}{\rmdefault}{\mddefault}{\updefault}{\color[rgb]{1,0,0}$\theta_{\alpha}$}%
}}}}
\put(3040,-7475){\makebox(0,0)[lb]{\smash{{\SetFigFont{8}{9.6}{\rmdefault}{\mddefault}{\updefault}{\color[rgb]{1,0,0}$\alpha$}%
}}}}
\put(401,-6708){\makebox(0,0)[lb]{\smash{{\SetFigFont{8}{9.6}{\rmdefault}{\mddefault}{\updefault}{\color[rgb]{.56,0,0}$\alpha_1$}%
}}}}
\put(2027,-6524){\makebox(0,0)[lb]{\smash{{\SetFigFont{8}{9.6}{\rmdefault}{\mddefault}{\updefault}{\color[rgb]{0,0,.69}$\beta$}%
}}}}
\put(2273,-6248){\makebox(0,0)[lb]{\smash{{\SetFigFont{8}{9.6}{\rmdefault}{\mddefault}{\updefault}{\color[rgb]{0,0,.69}electron}%
}}}}
\put(3408,-7383){\makebox(0,0)[lb]{\smash{{\SetFigFont{8}{9.6}{\rmdefault}{\mddefault}{\updefault}{\color[rgb]{1,0,0}emergent}%
}}}}
\put(3408,-7567){\makebox(0,0)[lb]{\smash{{\SetFigFont{8}{9.6}{\rmdefault}{\mddefault}{\updefault}{\color[rgb]{1,0,0}photon}%
}}}}
\put( 31,-6248){\makebox(0,0)[lb]{\smash{{\SetFigFont{8}{9.6}{\rmdefault}{\mddefault}{\updefault}{\color[rgb]{.56,0,0}incident}%
}}}}
\put( 31,-6432){\makebox(0,0)[lb]{\smash{{\SetFigFont{8}{9.6}{\rmdefault}{\mddefault}{\updefault}{\color[rgb]{.56,0,0}photon}%
}}}}
\put(1444,-7506){\makebox(0,0)[lb]{\smash{{\SetFigFont{8}{9.6}{\rmdefault}{\mddefault}{\updefault}{\color[rgb]{0,0,0}$\theta_1$}%
}}}}
\put(1751,-7537){\makebox(0,0)[lb]{\smash{{\SetFigFont{8}{9.6}{\rmdefault}{\mddefault}{\updefault}{\color[rgb]{0,0,0}$\theta$}%
}}}}
\put(1505,-7721){\makebox(0,0)[lb]{\smash{{\SetFigFont{8}{9.6}{\rmdefault}{\mddefault}{\updefault}{\color[rgb]{0,0,0}$\chi$}%
}}}}
\end{picture}%
\caption{Angles involved in the scattering process as viewed in the blob frame}
\label{fig_angles}
\end{figure}
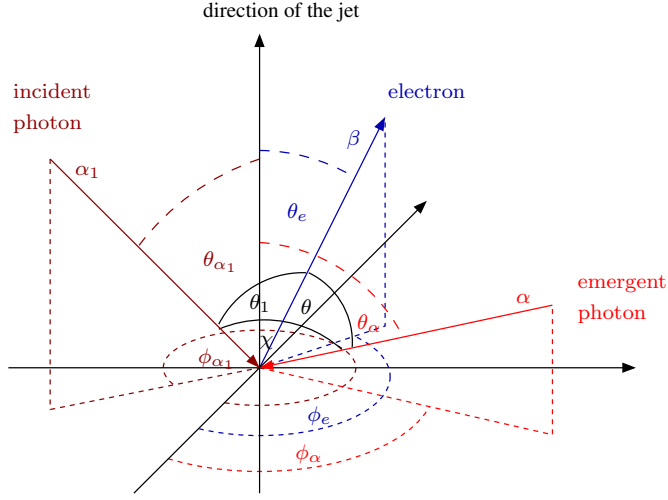

In this section we explain the calculation of the spectrum arising from the isotropically distributed, monoenergetic 
photons of energy $\alpha_1^*$, that have been Compton-scattered by the homogeneous electrons in the moving blob.
In the remainder of this work all quantities in the electron rest frame will be primed, while quantities in the blob frame are unprimed 
and quantities in the observer's frame are annotated by a star.
The photon and electron energies are given in units of the electron rest energy $m_ec^2$.
Figure \ref{fig_angles} illustrates the use of relative and absolute angles in the restframe of the blob, the blob frame.
The differential number of Compton scattered photons of energy $\alpha$ in the blob frame given by 
\begin{equation}
 \frac{\mathrm{d}n}{\mathrm{d}t\mathrm{d}\Omega_{\alpha}\mathrm{d}\alpha}=
\int \mathrm{d}\gamma\int \mathrm{d}\Omega_e\int \mathrm{d}\alpha_1 \int \mathrm{d}\Omega_{\alpha_1} n_{ph}(\alpha_1,\Omega_{\alpha_1}) n_e(\gamma)\ \sigma(\alpha_1,\Omega_{\alpha_1},\gamma,\Omega_e;\alpha,\Omega_{\alpha})
\end{equation}
whereas $n_{ph}$ is the incident photon distribution, $n_e$ the isotropic electron distribution and $\sigma$ the full Klein-Nishina cross section. 
We evaluate this expression in several steps:

Firstly, the resulting photon distribution in the blob frame is analytically calculated when a photon of energy $\alpha_1$ (emerging from $\Omega_{\alpha_1}$)
Compton-scatters an electron of energy $\gamma$ (emerging from $\Omega_e$). 
\begin{equation}
n_{\gamma,\Omega_e}^{(1)}(\alpha,\Omega_{\alpha})=\int \mathrm{d}\alpha_1 \int \mathrm{d}\Omega_{\alpha_1}  n_{ph}^{(1)}(\alpha_1,\Omega_{\alpha_1})\ \sigma(\alpha_1,\Omega_{\alpha_1},\gamma,\Omega_e;\alpha,\Omega_{\alpha})
\label{eq_onephoton}
\end{equation}
The incident photon distribution is given by
\begin{equation}
n_{ph}^{(1)}(\alpha_1,\Omega_{\alpha_1})=\frac{n_{ph}}{4\pi}\ \delta(\Omega_{\alpha_1}-\Omega_{\alpha_1}^0)\  \delta(\alpha_1-\alpha_1^0)
\end{equation}
and we assume the incident photons emerging from $\theta_{\alpha_1}^0=0$. $\delta$ denotes the Dirac delta function.
We use the full Klein-Nishina cross section in the electron rest frame is given by \citep{Jones1968}
\begin{eqnarray}
\frac{\mathrm{d}\sigma}{\mathrm{d}\alpha \ \mathrm{d}cos\chi'\ \mathrm{d}\phi'}&=&\frac{r_0^2 (1+\cos^2\chi')}{2\left[1+\alpha_1'(1-\cos\chi')\right]^2}\left[1+\frac{\alpha_1'^2(1-\cos\chi')^2}{(1+\cos\chi'^2)\left[1+\alpha_1'(1-\cos\chi')\right]}\right]\nonumber \\
&& \times\ \delta(\alpha'-\frac{\alpha_1'}{1+\alpha_1'(1-\cos\chi')}),
\end{eqnarray}
where $r_0=e^2/mc^2$ and $\chi'$ denotes the scattering angle in the electron rest frame. We transform the Klein-Nishina cross section into the blob frame by
\begin{equation}
\frac{\mathrm{d}\sigma}{\mathrm{d}\alpha\ \mathrm{d}\Omega_{\alpha}}=\frac{\mathrm{d}\sigma}{\mathrm{d}\alpha' \ \mathrm{d}\cos\chi'\ \mathrm{d}\phi'}\ \frac{\mathrm{d}\cos\chi'\ \mathrm{d}\phi'}{\mathrm{d}\cos\chi\ \mathrm{d}\phi}\ \frac{\mathrm{d}\cos\chi\ \mathrm{d}\phi}{\mathrm{d}\Omega_{\alpha}}\ \frac{\mathrm{d}\alpha'}{\mathrm{d}\alpha}.
\end{equation}
The transformations between relative 
($\theta$, $\theta_1$, $\chi$) and absolute coordinates ($\theta_{\alpha}$, $\phi_{\alpha}$) (cf. Figure \ref{fig_angles}) are given by
\begin{eqnarray}
\cos\chi&=&\cos\theta_{\alpha}\nonumber\\
\cos\theta_1&=&\cos\theta_e\\
\cos\theta&=&\cos\theta_{\alpha} \cos\theta_e+\sin\theta_{\alpha}\sin\theta_e \cos(\phi_{\alpha}-\phi_e)\nonumber
\end{eqnarray}
while the transformations between electron rest and blob frame are as follows:
 \begin{eqnarray}
 \frac{\mathrm{d}\cos\chi'}{\mathrm{d}\cos\chi}&=&\frac{1}{\gamma^2 (1+\beta\cos\theta)(1+\beta\cos\theta_1)}\nonumber\\
 \frac{\mathrm{d}\phi'}{\mathrm{d}\phi}&=&\frac{1+\beta\cos\theta_1}{1+\beta\cos\theta}\\
 \frac{\mathrm{d}\alpha'}{\mathrm{d}\alpha}&=&\gamma (1+\beta\cos\theta)\nonumber
 \end{eqnarray}

Secondly, we calculate the emergent photon distribution when a photon Compton scatters a homogeneous electron distribution.
Hence, because of the homogeneous electron distribution $n_e(\gamma)$, the resulting analytic term from equation (\ref{eq_onephoton}) is numerically integrated 
over the solid angle of the electrons $\Omega_e$ in the blob frame. 
\begin{equation}
n_{\gamma}^{(1)}(\alpha,\Omega_{\alpha})=\int \mathrm{d}\Omega_e\ n_{\gamma,\Omega_e}^{(1)} (\alpha,\Omega_{\alpha})
\label{eq_electronangle}
\end{equation}
We checked the accuracy of the results by revisiting the properties of Compton-scattering, e.g. forward scattering for only $\alpha=\alpha_1$ or the maximal energy gained by scattering.

Thirdly, we assume an incident monoenergetic isotropic photon distribution in the observer's frame $n_{ph}^*(\alpha_1^*,\Omega_{\alpha_1}^*)=n_{ph,0}^*$ and transform it into the blob frame
\begin{equation}
n_{ph}(\alpha_1,\Omega_{\alpha_1}) = \frac{n^*_{ph,0}}{\Gamma^3 (1-B\cos\theta_{\alpha_1})^3} \ \delta \left(\alpha_1-\frac{{\alpha_1^*}_0}{\Gamma (1-B\cos\theta_{\alpha_1})}\right),
\end{equation}
This allows us to integrate the obtained result in equation (\ref{eq_electronangle}) over the incident photon distribution $n_{ph}(\alpha_1,\Omega_{\alpha_1})$. 
For that purpose the emergent solid angle $\Omega_{\alpha}$ is rotated with respect to the considered incident photon direction $\Omega_{\alpha_1}$, i.e. the new  $\Omega_{\alpha}$ is given by $\hat{R}_{\Omega_{\alpha_1}}\Omega_{\alpha}$.
$\hat{R}_{\Omega_{\alpha_1}}$ represents the rotation operator with respect to $\Omega_{\alpha_1}$.
\begin{equation}
n_{\gamma}(\alpha,\Omega_{\alpha})=\int \mathrm{d}\alpha_1\int \mathrm{d}\Omega_{\alpha_1} n_{ph} (\alpha_1,\Omega_{\alpha_1}) \ n_{\gamma}^{(1)} (\alpha, \hat{R}_{\Omega_{\alpha_1}} \Omega_{\alpha})
\label{eq_photons}
\end{equation}
The integration over all directions ($\Omega_{\alpha_1}$) and energies ($\alpha_1$) is done numerically.

Fourthly, we assume the energetic distribution of the electrons as $n_e(\gamma)=n_e^0\ \gamma^{-p}$ and consider energies ranging from $\gamma=1$ to $\gamma=\gamma_{max}=10^{12}$.
In order to integrate over the electron energies each number of electrons of energy $\gamma$ is multiplied by the respective calculated differential number of the Compton-scattered photons in the third step (cf. equation (\ref{eq_photons})). 
The obtained values are summed up for all energies $\gamma$.
\begin{equation}
n(\alpha,\Omega_{\alpha})=\int_1^{\gamma_{max}} \mathrm{d}\gamma\ n_e(\gamma)\ n_{\gamma}(\alpha,\Omega_{\alpha})
\end{equation}
Finally, we transform the resulting differential photon distribution expressed in the blob frame into the observer's frame using
\begin{equation}
\frac{\mathrm{d}n^*}{\mathrm{d}t^*\ \mathrm{d}\alpha^*\ \mathrm{d}\Omega_{\alpha}^*}=\frac{\mathrm{d}n}{\mathrm{d}t\ \mathrm{d}\alpha\ \mathrm{d}\Omega_{\alpha}}\ \Gamma(1-B\cos\theta_{\alpha})^2.
\end{equation}
The intensity is yielded by
\begin{equation}
I^*\ =\ I\ \Gamma^3 (1-B\cos\theta_{\alpha})^3\ =\ \alpha\ \frac{\mathrm{d}n}{\mathrm{d}t\ \mathrm{d}\alpha\ \mathrm{d}\Omega_{\alpha}}\ \frac{V}{d^2}\ \Gamma^3 (1-B\cos\theta_{\alpha})^3
\end{equation}
where $V$ is the volume of the blob and $d^2$ its area perpendicular to the moving direction of the blob.

\section{Results and discussion}
\label{sec_results}
The characteristic spectrum of the Compton-scattered photons is a broadened peak if we assume an isotropic electron distribution in the blob frame. 
When plotting the spectrum on a double logarithmic scale graph, two characteristic areas show up. 
These are the regime of elastic scattering, i.e. the Thomson regime, and the regime of inelastic scattering, i.e. the Klein-Nishina regime. 
Defining the electron distribution by a spectral index $p$ the differential number of photons in the Thomson regime depends on $\alpha^{-(p-1)/2}$. 
In the Klein-Nishina regime the spectral index is characterized by a higher spectral index.

\subsection{Blob frame}
Choosing an electron distribution of $n_e(\gamma)=\gamma^{-2.2}$, the plot of the differential number of scattered photons displays the transition from the Thomson (elastic scattering) to the Klein-Nishina regime (inelastic scattering) that is displayed as a sharp break in Figure \ref{fig_comp_dermer}. 
The upper limit of the Thomson regime depends on the initial energy in the observer's frame because depending
on the incident angle of the external photons they loose or win energy due to the change of the frame of reference.
The maximal electron energy $\gamma_{T,max}$ that allows scattering in the Thomson regime determines this upper limit: $\alpha_{1,min} \gamma_{T,max} <1$.
Hence, the maximal photon energy that results from Thomson scattering is defined by the maximal electron energy, i.e. $\alpha_{T,max}=4 \alpha_{1,min} \gamma_{T,max}^2$.
The higher $\Gamma$, the more the incident photon distribution is peaked, the higher the maximum energy $\alpha_{1,max}=2\Gamma \alpha_1^*$
and the lower $\gamma_{T,max}$ and the Thomson limit.

\subsection{Observer's frame}
\begin{figure}[htb]
\centering
\includegraphics[width=11cm]{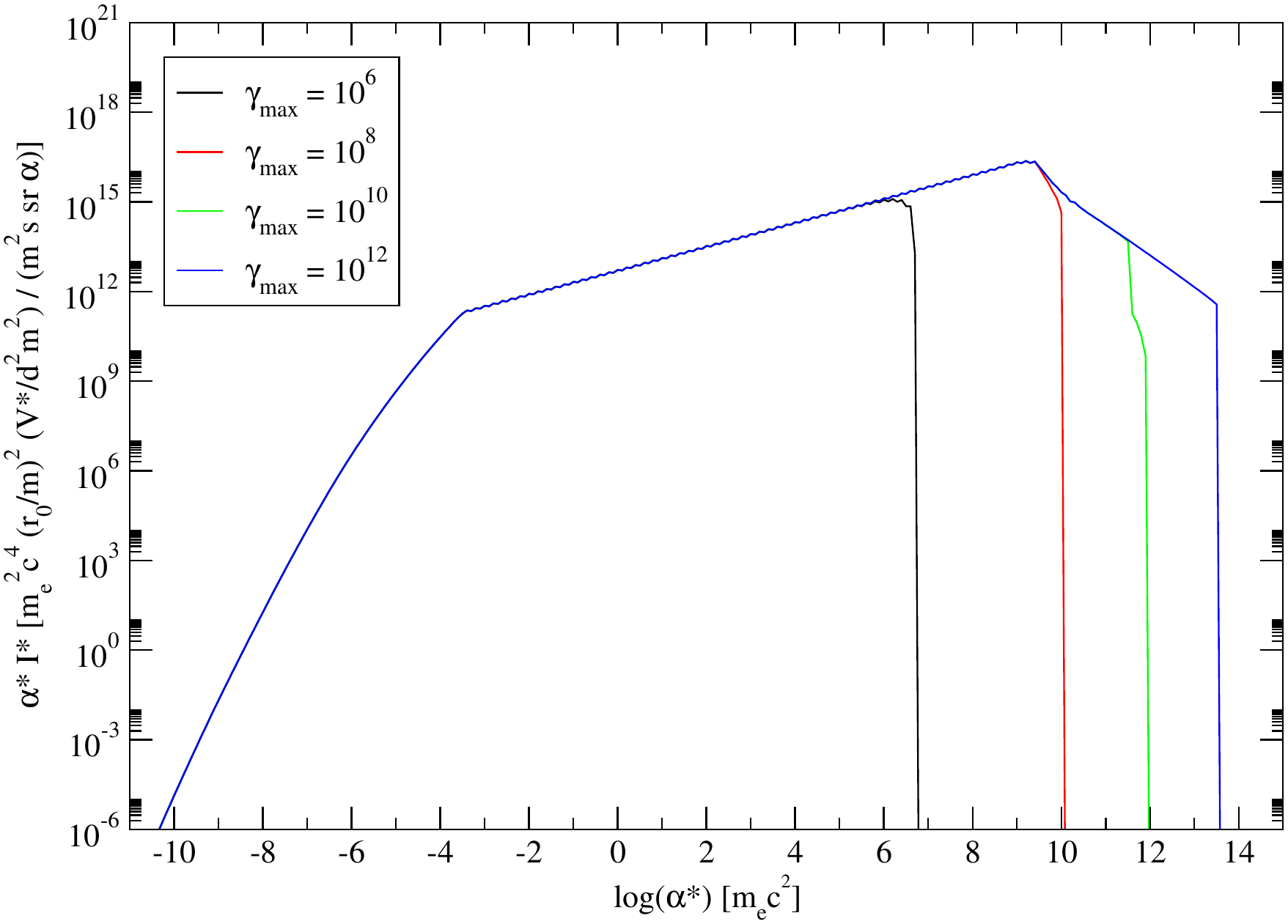}
\caption{Comparison of the gamma ray emissivities assuming different maximal electron energies $\gamma_{max}=10^6$, $10^8$, $10^{10}$, $10^{12}$. 
Emissions are presented in the observer frame assuming a spectral index of the electron distribution of $p=2.2$, $n_e^0=1$, $n_{ph}=1$ and an incident photon energy of $\alpha^*_1=10^{-9}$.
Emissions are calculated in the observer frame and correspond to a viewing angle of $0^{\circ}$ if $\Gamma=20$.}
\label{fig_energies}
\end{figure}
\begin{figure}[htb]
\centering
\includegraphics[width=11cm]{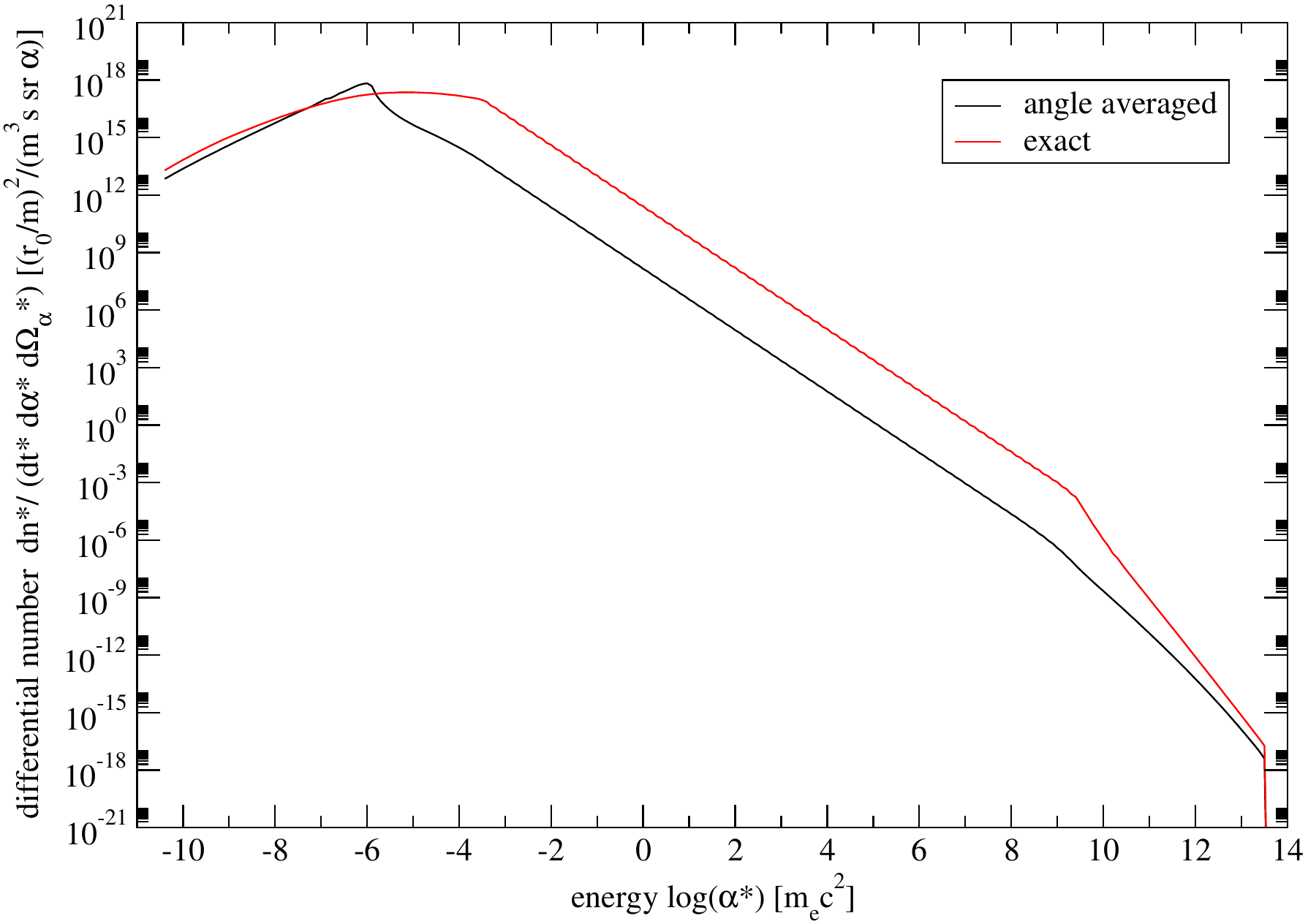}
\caption{Comparison of the gamma ray emissivities using angle resolved and angle averaged calculations.
Emissions are presented in the observer frame assuming a spectral index of the electron distribution of $p=2.2$, $n_e^0=1$, $n_{ph}=1$ and an incident photon energy of $\alpha^*_1=10^{-9}$.
Emissions are calculated in the observer frame and correspond to a viewing angle of $0^{\circ}$ if $\Gamma=20$.}
\label{fig_comp_georg}
\end{figure}
Boosting the resulting spectrum into the observer's frame enhances the differential number of photons emerging at $\theta^*_{\alpha}\rightarrow \pi$ and decreases those emerging at $\theta^*_{\alpha}\rightarrow 0$.
The main characteristics of the spectrum - in particular the break - remain.
In Figure \ref{fig_energies} the development of the break is shown by choosing different maximal electron energies of $\gamma_{max}=10^{6}$, $10^8$, $10^{10}$, $10^{12}$. The higher the electron energy, the higher the maximal energy of the scattered photons. In the Klein-Nishina regime photons are scattered to energies up to $\alpha^*_{KN,max}=2\Gamma\gamma_{max}$ while in the Thomson regime up to $\alpha^*_{T,max}=2\Gamma\ 4 \alpha_{1,min} \gamma_{T,max}^2$. The break gradually appears as scattering is dominated by the Klein-Nishina regime.
It turns out that this break is sharper than in \citet{Georganopoulos2001}. 
As we do not use any approximations (e.g. head on approximation, outgoing photons are directed along the direction of the scattering electrons, i.e. $\Omega_{\alpha}=\Omega_e$) when calculating the spectrum the sharper break is due to the angle averaged solution of \citet{Georganopoulos2001} (introduced by \citet{Jones1968}).
By taking the average of all emergent photon angles in the blob frame we studied how angle averaging effects the shape of the break (cf. Figure \ref{fig_comp_georg}). The break becomes less sharp and is similar to the shape of the break in \citet{Georganopoulos2001}. The number of photons of the angle averaged solution is lower than those of the exact calculation due to the decreasing photon number at lower emerging angle $\theta^*_{\alpha}$. 

\subsection{Comparison with \citet{Dermer1995}}

\begin{figure}[htb]
\centering
\includegraphics[width=11cm]{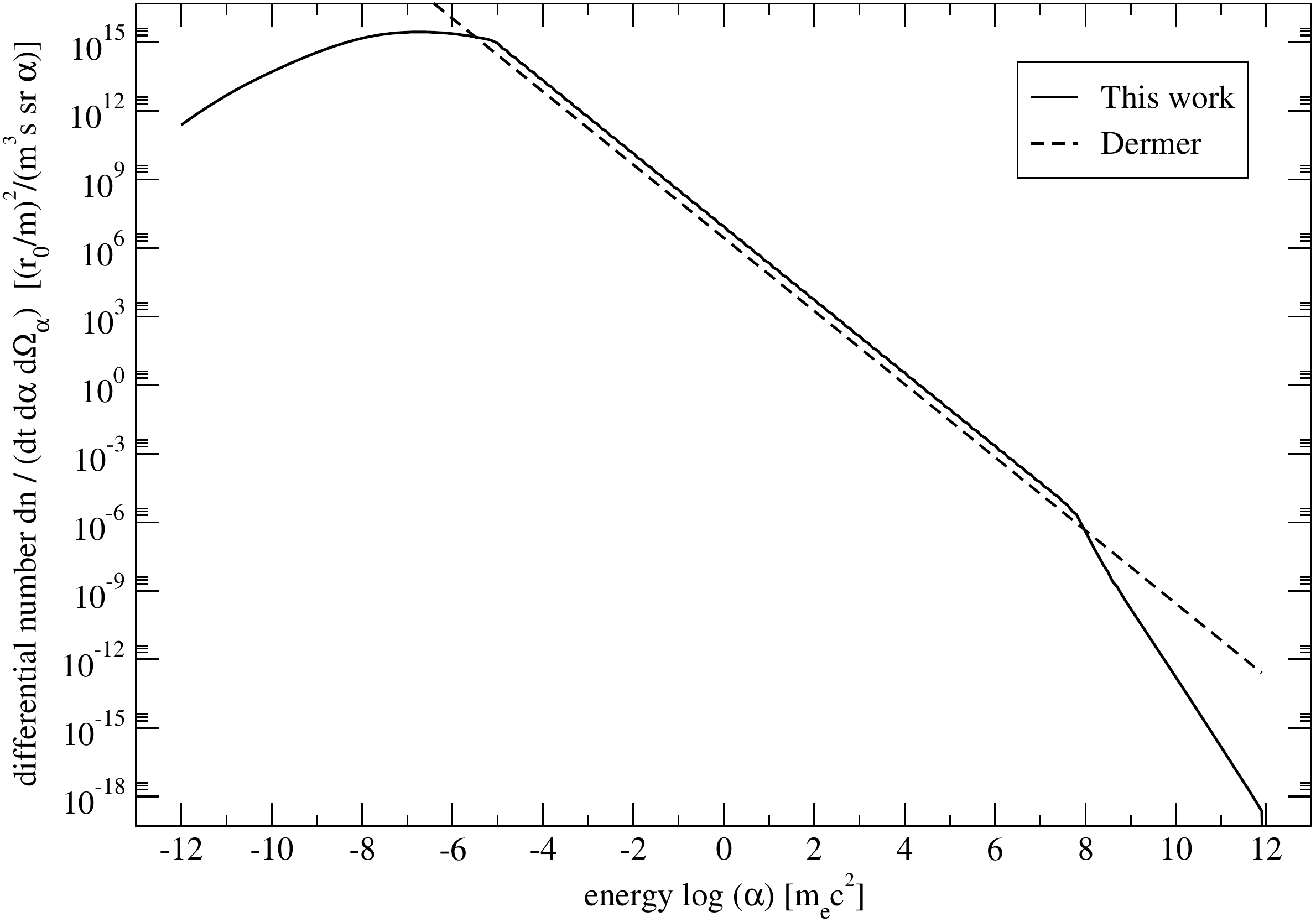}
\caption{Comparison between the gamma ray emissivities using the full Klein-Nishina cross section and Dermer's approximation. Emissions are calculated in the blob frame and correspond to a viewing angle of $0^{\circ}$ if $\Gamma=20$. The spectral index for the electron distribution runs $p=2.2$, $n_e^0=1$, $n_{ph}=1$, $1\leq \gamma\leq 10^{12}$.}
\label{fig_comp_dermer}
\end{figure}
\begin{figure}[htb]
\centering
\includegraphics[width=12cm]{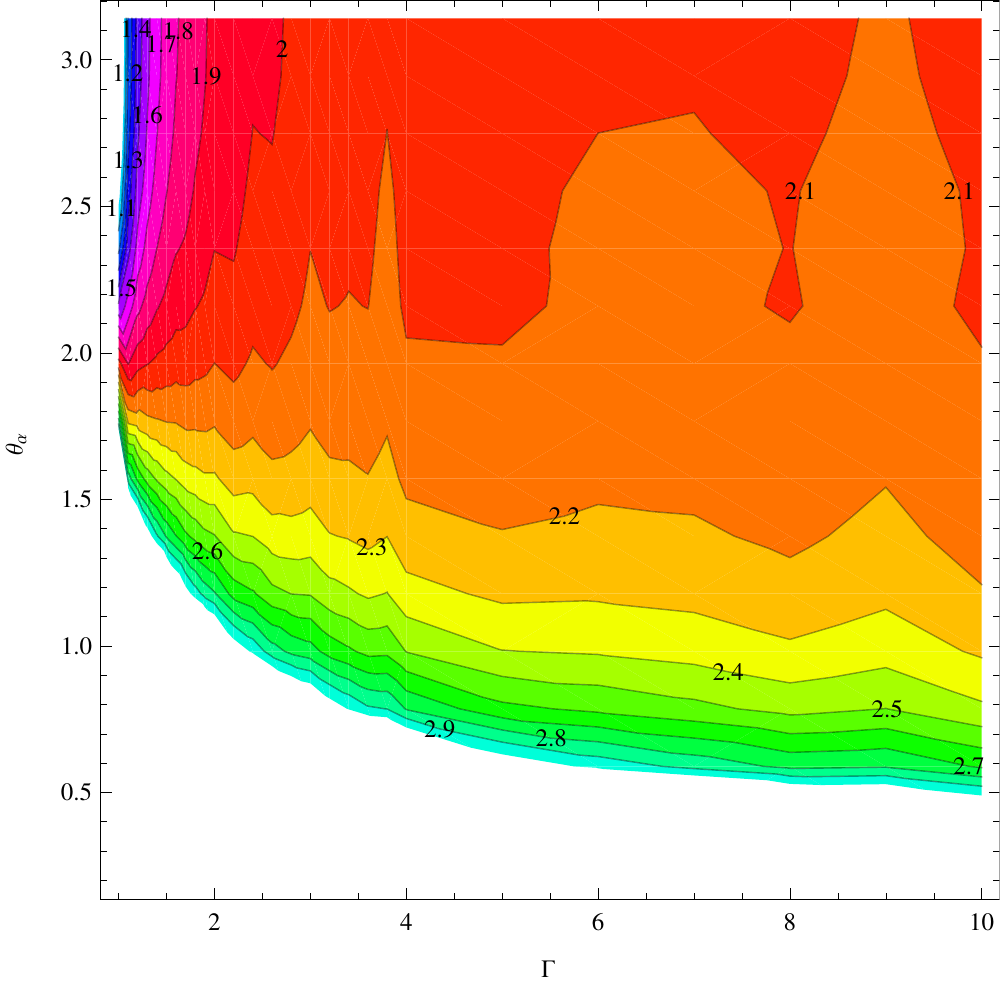}
\caption{Relative difference $\epsilon$ of the differential number of scattered photons between this exact calculation and \citet{Dermer1995}, dependent on the emergent angle $\theta$ and the gamma factor of the blob $\Gamma$.
The lines are contour lines and show together with the coulors the value of the relative difference.}
\label{fig_valid_dermer}
\end{figure}

In the following section we compare our results to those of  \citet{Dermer1995} with respect to the regime of elastic scattering (Thomson regime).
We observe the differential scattered photon number to be larger than predicted by \citet{Dermer1995} (cf. Figure \ref{fig_comp_dermer}).
Dermer considers only the emergent photons that are scattered to the maximal energy $\alpha_{max}=4\alpha_1 \gamma^2$. 
This fact explains the observed difference since in our work all photons with energies ranging from the initial photon energy to the maximum energy $\alpha_{max}$ are also considered. We obtain Dermer's result if we only take into account the photons that are scattered to the maximum energy.
The distribution of scattered photons in the observer's frame as well as the difference to Dermer highly depends on the observing angle and the velocity of the blob.
Depending on the emergent angle $\theta_{\alpha}$ and the gamma factor of the blob $\Gamma$ the relative difference $\epsilon$ of the differential number of scattered photons between our exact calculation and \citet{Dermer1995} is shown in Figure \ref{fig_valid_dermer} as a contour plot. The mean value for all relative differences for $2.00\leq\epsilon\leq2.12$ is given by $\epsilon_{lim}=2.09$ and hence the correcting factor runs $f_{corr}=1+\epsilon_{lim}=3.09$. The limits for the gamma factor of the blob $\Gamma$ and the emergent angle $\theta_{\alpha}$ can be deduced from the values in Figure \ref{fig_valid_dermer}.

We find that for higher emergent angles $\theta_{\alpha}>2.09\ \mathrm{rad}$ and higher blob velocities $\Gamma>2.8$ our differential number of scattered photons is higher by a factor of $f_{corr}=3.09$ than calculated in \citet{Dermer1995}. 
Hence, Dermer's approximations can be used for high Doppler factors and small observing angles, i.e. $\theta_{\alpha}=\pi$. 
But for higher observing angles and smaller Doppler factors the deviation increases and
the Klein-Nishina cross section and the exact incident photon distribution must be considered.
Considering the boost factor that is needed to transform the differential number of scattered photons from the blob frame into the observer's frame, we computed a lower  boost factor than derived in \citet{Dermer1995}. 
Analogously to \citet{Dermer1995} the boosting depends on the spectral index $p$ of the electron distribution. 
If $a_{\alpha}$ and $a_{\theta_{\alpha}}$ describe the exponents of the Dopplerfactor of the respective quantities, the found boost factor goes as $D^{3+a_{\alpha}+1+a_{\theta_{\alpha}}}$.
$a_{\alpha}$ coincidences with Dermer's $a=(p-1)/2$ and $a_{\theta_{\alpha}}$ reaches a smaller value.
This difference can be explained by Dermer's use of the head on approximation. 
Dermer uses this approximation for all scattering processes but actually it is not valid for low energetic electrons. 
Furthermore the number of low energetic electrons coincides with the decrease of the deviation when smaller spectral indices of the electron distribution are considered.

\subsection{Conclusions}
Comparing our work to \citet{Georganopoulos2001} we did not use the head on approximation as well as the calculated scattered spectrum by \citet{Jones1968} that averages over all angles of the emergent photons. 
In \citet{Georganopoulos2001} the only angle dependent quantity is the Doppler factor while we computed directly the angle dependent scattered spectrum. While the rough shape and peak energies coincidences with \citet{Georganopoulos2001} we find as already mentioned a smaller boost factor than \citet{Dermer1995} and \citet{Georganopoulos2001}.
Due to the exact calculation the break in our spectrum is much sharper than in \citet{Georganopoulos2001}. But taking the average of all angles of the emergent photons softens the break and approaches the results of \citet{Georganopoulos2001}.

Summing up, in contrast to \citet{Dermer1995} we observe a higher differential number of scattered photons in the regime of elastic scattering. 
Furthermore a lower boost factor was computed because we considered the exact distribution of the incident photon field. 
Despite this difference Dermer's approximation is applicable for high Doppler factors and small observing angles.
But for high inclination angles (e.g. M87, \citealt{Giovannini2010}) or small Doppler factors (e.g. BL Lacertae 3C371.0 or RG 3C84, \citealt{Hovatta2009}) the calculations of this work should be used.
Furthermore for high electronic spectral indices Dermer's boost factor overvalues and a too large number of scattered photons is yielded. 

\section*{Acknowledgements}
We like to thank the anonymous referees for their useful comments on a previous version of this paper.

\end{document}